\documentclass[amsmath,nofootinbib,twocolumn,superscriptaddress]{revtex4-1}
\pdfoutput=1
\usepackage[colorlinks=true,citecolor=blue,urlcolor=blue]{hyperref}
\usepackage{aas_macros,graphicx,marvosym,subfigure,amssymb,amsmath}
\usepackage{comment}
\usepackage{slashed}

\newcommand{\pd}{\,\partial}

\newcommand{\NHN}[1]{\bar{#1}}

\begin{document}

\title{Transient Instability of Rapidly Rotating Black Holes}

\author{Samuel E. Gralla}
\affiliation{Department of Physics, University of Arizona, Tucson, Arizona 85721, USA}

\author{Aaron Zimmerman}
\affiliation{Canadian Institute for Theoretical Astrophysics, 60 St. George Street, Toronto, Ontario, M5S 3H8, Canada}

\author{Peter Zimmerman}
\affiliation{Department of Physics, University of Arizona, Tucson, Arizona 85721, USA}

\begin{abstract}
We analytically study the linear response of a near-extremal Kerr black hole to external scalar, electromagnetic, and gravitational field perturbations.  We show that the energy density, electromagnetic field strength, and tidal force experienced by infalling observers exhibit transient growth near the horizon.  The growth lasts arbitrarily long in the extremal limit, reproducing the horizon instability of extremal Kerr.  We explain these results in terms of near-horizon geometry and discuss potential astrophysical implications.
\end{abstract}

\maketitle

\section{Introduction}

Black holes play a central role in modern theoretical physics and astrophysics.  A special case of considerable interest is the ``extremal'' limit of vanishing surface gravity.  Higher-dimensional extremal black holes play an important role in string theory \cite{Strominger1996}.  For the Kerr black holes of our Universe, extremal means maximally spinning.  There has been much recent interest in rapidly spinning, near-extremal black holes in light of their enhanced symmetries \cite{bardeen-horowitz1999,kerrCFT}, conjectured holographic duality \cite{kerrCFT}, unusual dynamics \cite{Yang:2014tla}, analytic tractability \cite{TeukolskyPress1974,porfyriadis-strominger2014,hadar-porfyriadis-strominger2014,lupsasca-rodriguez-strominger2014,zhang-yang-lehner2014,lupsasca-rodriguez2015,hadar-porfyriadis-strominger2015,GPW15,gralla-lupsasca-strominger2016,compere-oliveri2016,porfyriadis-shi-strominger2016}, and unique observational signatures \cite{Andersson2000,Glampedakis2001,Yang:2013uba,Gralla:2016qfw,Burko:2016sfi}.

In 2010 Aretakis discovered that extremal black holes are unstable \cite{Aretakis:2010gd,Lucietti2012,Aretakis:2012ei}.  He showed that sufficiently high-order derivatives on the event horizon grow unboundedly with time.  Since no physical object can be \textit{exactly} extremal, the physical implications of the instability rest on generalization to near-extremal black holes.  This was done for spherically symmetric \textit{nonlinear} perturbations of charged (Reissner-Nordstr\"om) black holes in beautiful numerical work by Murata, Reall, and Tanahashi \cite{Murata:2013daa}, who found transient growth on the horizon. This growth lasts arbitrarily long in the extremal limit, recovering the unbounded growth of the extremal instability.

We provide similar results for the astrophysical Kerr black hole.  Our calculations are limited to linearized theory, but have the advantages of being both analytical and covering the nonaxisymmetric modes, which dominate the extremal instability \cite{Casals:2016mel}.  We show that the instability is associated with a family of ``zero damped'' quasinormal modes~\cite{Hod2008a,Yang:2012pj,Yang:2013uba}, which we call near-horizon modes.  Generic initial data produces a coherent excitation that gives rise to transient growth near the horizon. Increasing the spin shrinks the region of growth while lengthening the growth time, recovering the Aretakis instability---unbounded growth only precisely on the horizon---in the extremal limit.

The above discussion implicitly assumes that the extremal limit is taken in one of the usual coordinate systems (such ingoing Kerr coordinates), which produces the metric known as extremal Kerr.  An alternative extremal limit adapted to near-horizon observers produces a different metric known as near-horizon extremal Kerr (NHEK) \cite{bardeen-horowitz1999}.  The singular relationship between the limits means that near-horizon excitations are  singular to far-horizon observers (and vice versa).  The instability is in effect the statement that near and far dynamics do not completely decouple in the extremal limit, making singular behavior unavoidable.

Among the physical quantities that grow in response to external perturbations are energy densities, electromagnetic field strengths, and tidal forces measured by infalling observers.  The large observed energy density is analogous to the high-energy particle collisions that can be produced with finely tuned initial data \cite{PSK75,Banados:2009pr}, except that here no tuning is required.  The growth of electromagnetic fields means that rapidly spinning black holes act to amplify generic external fields, a fact with potential observational consequences for radiation from charged particles.  Perturbing tidal forces provide a small enhancement of the black hole's own tidal fields, and this amplification may encourage the development of gravitational turbulence \cite{Yang:2014tla}.  Further study is required to explore these potential consequences of the transient instability.

In what follows we derive the transient instability, discuss it in terms of near-horizon geometry, and elaborate on the physical implications. 
Geometric units $G = c = 1$ are used throughout.

\section{Near-Horizon Quasinormal Mode Response}\label{sec:GreenFunc}

We investigate the perturbations of Kerr black holes using ingoing Kerr coordinates $x^\mu = (v, r, \theta, \varphi)$ \cite{BoyerLindquist1967} around a black hole of mass $M$ and spin parameter $a$.
The outer and inner horizons lie at $r_\pm  = M \pm \sqrt{M^2 -a^2}$, respectively, and the outer horizon rotates at the horizon frequency $\Omega_H = a/(2Mr_+)$.

\subsection{Mode decomposition}
Teukolsky \cite{Teukolsky1973} showed that in suitable tetrads on Kerr certain perturbed Newman-Penrose \cite{NewmanPenrose1962} scalars obey decoupled, separable equations.  These scalars contain all the radiative information about the corresponding perturbations \cite{Wald:1973jmp,Cohen:1974,Chrzanowski:1975wv,Wald:1978vm}.  We work with the scalars $\Omega_s$ defined in \cite{TeukolskyPress1974} for $s=0, \pm 1, \pm 2$ corresponding to scalar, electromagnetic, and gravitational fields, respectively (see Appendix \ref{sec:tetrad} for details).  For source-free perturbations these obey a second-order linear partial differential equation $L_s[\Omega_s]=0$.  We consider the Green function $G$ for this operator,
\begin{align}
L_s [G] = \delta^{(4)}(x^\mu - x^{\mu}{}').
\end{align}
The equation separates under mode decomposition and a Laplace transform,
\begin{align}
\label{eq:IngoingGF}
&G(x^\mu,x^\mu{}')   = \frac{1}{2\pi}  \sum_{\ell =|s|}^{\infty} \sum_{m=-\ell}^{\ell}e^{i m(\varphi - \varphi')} \\ &\times 
\int_{-\infty+ ic}^{\infty +ic} e^{-i \omega(v-v')}  {}_sS_{\ell m\omega}(\theta) {}_sS^*_{\ell m\omega}(\theta') 
\tilde g_{\ell m \omega}(r, r') \, d \omega,\nonumber
\end{align}
where ${}_s S_{\ell m\omega}(\theta)$ are spin-weighted spheroidal harmonics \cite{Teukolsky1973} and $c$ is a positive constant. 
The radial function $\tilde{g}_{\ell m \omega}$ satisfies the ordinary differential equation (A1) of \cite{TeukolskyPress1974} with $\delta(r-r')$ on the right-hand side.  The causal solution is constructed from homogeneous solutions $R^{\rm in}$ with no incoming radiation from the horizon and $R^{\rm up}$ with no incoming radiation from infinity via
\begin{align}\label{eq:radialg}
\tilde g_{\ell m \omega}(r, r') & = \frac{R^{\rm in}(r_<) R^{\rm up} (r_>)}{\mathcal W} \,,
\end{align}
where  $r_> = \max (r, r')$, $r_< = \min (r, r')$. 
Here $\mathcal{W} = 
\Delta^{s+1} e^{-2 i \omega r_* } \left(R^{\rm in} \pd_r R^{\rm up}  - R^{\rm up}  \pd_r R^{\rm in}\right)$ with $\Delta = (r-r_-)(r-r_+)$ and where $r_*$ is the tortoise coordinate \cite{Teukolsky1973}.
We restrict to nonaxisymmetric modes, $m \neq 0$.\footnote{It is possible, but cumbersome, to treat axisymmetric and nonaxisymmetric modes in a unified notation \cite{Casals:2016mel}.  We are confident that the nonaxisymmetric modes are dominant since these dominate the extremal instability \cite{Casals:2016mel}.}   With a convenient choice of overall normalization, the up solution satisfies \cite{TeukolskyPress1974}
\begin{align}\label{Ruplarger}
R^{\rm up}(r) \sim \frac{e^{2i\omega r_*}}{r},\quad r \rightarrow \infty .
\end{align}
Similarly, we normalize the in solution such that \cite{TeukolskyPress1974}
\begin{align}
R^{\rm in} & \sim \left \{
\begin{array}{ll}
Z^{\rm out} r^{-1} e^{2 i \omega r_*} + Z^{\rm in} r^{-2s-1}  & r \to \infty \\
1 & r \to r_+ \\
\end{array} \right. \,,
\end{align}
where $Z^{\rm in}$ and $Z^{\rm out}$ may be determined by solving the radial equation.  In terms of these definitions we have
\begin{align}\label{Wronky}
\mathcal{W} = 2 i \omega Z^{\rm in}.
\end{align}

\subsection{Near-extremal case}

To study the near-extremal regime we now introduce dimensionless quantities 
\begin{align}\label{sigmax}
\sigma = \frac{r_+-r_-}{r_+}, \qquad x = \frac{r-r_+}{r_+},
\end{align}
defined so that $x=0$ is the horizon and $\sigma \rightarrow 0$ is the extremal limit.  Teukolsky and Press~\cite{TeukolskyPress1974} used matched asymptotic expansions to find analytic solutions valid for frequencies near the superradiant bound $(\omega - m \Omega_H)r_+ \ll 1$ in the near-extremal regime $\sigma \ll 1$.

The results needed here are the in solution near the horizon ($x \ll 1$),
\begin{align}\label{Rin}
R^{\rm in}(x) = {}_2 F_1(\alpha_+,\alpha_-, 1 + s - 2 i \NHN \omega; - x/\sigma)\,,
\end{align}
and the incident wave amplitude,
\begin{align}
Z^{\rm in}  = & \frac{ (- i m)^{-1/2 - s + i \delta - i m} \Gamma(-2 i \delta)\Gamma(1 - 2 i \delta)}{\Gamma(\alpha_-) \Gamma(1/2-s-i \delta -  i m)} 
\notag \\ & \times
\frac{\Gamma(1 + s -2i\NHN \omega)}{\Gamma(1/2 -i\delta + i m - 2 i \NHN \omega)} 
\sigma^{\alpha_+} +  (\delta \to - \delta)\,. \label{Zin}
\end{align}
The notation $(\delta \to - \delta)$ means to repeat the same terms with the sign of $\delta$ reversed.  Here we have defined
\begin{align}
\alpha_{\pm} & = 1/2 + s \pm i \delta -  i m  \,, \label{alpha} \\
\NHN \omega & =  \frac{ 2M(\omega -   m \Omega_H)}{ \sigma }
\end{align}
with
\begin{align}\label{delta2}
\delta^2 =  \frac{7m^2}{4} - (s+1/2)^2 - {}_s A_{\ell m}  \,.
\end{align}
Here ${}_s A_{\ell m}$ is the eigenvalue ${}_sA_{\ell m \omega}$ of \cite{Teukolsky1973} evaluated at $a=M$ and $\omega=1/(2M)$.\footnote{The eigenvalue ${}_sA_{\ell m \omega}$ is related to the eigenvalue ${}_sK_{\ell m \omega}$ of  \cite{porfyriadis-strominger2014,hadar-porfyriadis-strominger2014,hadar-porfyriadis-strominger2015,GPW15,Casals:2016mel} by ${}_s K_{\ell m \omega} = {}_sA_{\ell m \omega} + s(s+1) + a^2 \omega^2 $.}  
Equation \eqref{delta2} defines $\delta$ only up to sign, with \eqref{Rin} and \eqref{Zin}  invariant under $\delta \rightarrow -\delta$.  We choose the convention $\delta = \sqrt{\delta^2}$; i.e.~$\delta$ is positive when real and has positive imaginary part when imaginary.

The cases $\delta^2>0$ and $\delta^2<0$ generally give rise to qualitatively different behavior \cite{Yang:2012he,Yang:2012pj,Yang:2013uba,GPW15}.  We name these cases ``principal'' and ``supplementary'' following terminology used in the representation theory of $\mathsf{SL}(2,\mathbb R)$ \cite{Barut:1965,Balasubramanian:1998sn}, a group that appears as part of the near-horizon isometry group (Appendix \ref{sec:Millerites}).  For each $\ell$ and $m$ one can determine whether the mode is principal or supplementary by computing the eigenvalue ${}_s A_{\ell m}$ and checking the sign of \eqref{delta2}.  These occur for larger and smaller values of $m/\ell$, respectively, with the transition at $m/\ell \approx 0.74$ \cite{Yang:2012he,HodEikonal2012} for all values of $s$ in the large-$\ell$ limit.  
The principal modes are also connected to the near-horizon photon orbits of Kerr via the geometric correspondence between the large-$\ell$ quasionormal modes (QNMs) and unstable null orbits \cite{Yang:2012he,Hod:2012ax,Yang:2013uba}.  In another common notation \cite{porfyriadis-strominger2014,hadar-porfyriadis-strominger2014,hadar-porfyriadis-strominger2015,GPW15,Casals:2016mel} the principal and supplementary modes correspond to complex and real  conformal weight $h$, respectively.  Table~\ref{tab:conventions} summarizes the properties of and conventions for these modes.

\begin{table}
\caption{Properties of nonaxisymmetric near-extremal modes and the relationships between the $\delta$ notation used here (and in \cite{TeukolskyPress1974}) and the $h$ notation used in \cite{porfyriadis-strominger2014,hadar-porfyriadis-strominger2014,hadar-porfyriadis-strominger2015,GPW15,Casals:2016mel}.}
\begin{tabular}{|r|c|c|c|c|}
\hline
Principal  & $\delta^2 > 0$  & $h \in \mathbb{C}$ & $h = \tfrac{1}{2} + i \delta$ & $ m \gtrsim 0.74 \ell$ \\
Supplementary & \ $\delta^2< 0$ \ & \ $h \in \mathbb{R}$ \ & \ $h = \tfrac{1}{2} - i \delta$ \ & \ $ m \lesssim 0.74 \ell$ \ \\
\hline
\end{tabular}
\label{tab:conventions}
\end{table}

\subsection{Overtone sum}

In order to calculate the Green function in the time domain, we must resolve the inverse Laplace transform in Eq.~\eqref{eq:IngoingGF}.  
Doing so results in three terms: the contribution from the arcs at large $|\omega|$, a contribution from a branch cut extending from $\omega = 0$ along the negative imaginary axis, and a sum over the poles of the Green function.  We focus on this last term, which is the contribution to the response from the decaying resonances of the black hole, known as the QNMs~\cite{Berti2009,Kokkotas1999}, which dominate the response at intermediate times following the initial signal propagating on the light cone \cite{Leaver1986b}.

The QNM frequencies are the poles of $\tilde g_{\ell m \omega}$, which by \eqref{eq:radialg} and \eqref{Wronky} occur when $Z^{\rm in}$ vanishes.  
From Eq.~\eqref{Zin} the QNM resonance condition for near-horizon modes  \cite{Detweiler1980,Hod2008a} is thus 
\begin{align}\label{eq:resonanceC}
 (- i m \sigma)^{-2i\delta} & \frac{\Gamma(2i\delta)^2 \Gamma(\alpha_-)}{\Gamma(-2 i \delta)^2 \Gamma(\alpha_+)}\frac{\Gamma(1/2 -i\delta +  i m - 2 i \NHN \omega)}{\Gamma(1/2 + i\delta +  i m - 2  i \NHN \omega )} \notag \\
&\times 
 \frac{ \Gamma(1/2 - s - i \delta -  i m)}{ \Gamma(1/2 - s +i \delta -  i m)} = 1.
\end{align}
For supplementary modes $\delta^2<0$, the quantity $(- i m \sigma)^{-2 i \delta}= O(\sigma^{2 |\delta|})$ is perturbatively small in $\sigma$ and must be compensated by a divergence in the multiplying factors in order to satisfy \eqref{eq:resonanceC}.  Noting that the gamma function has simple poles at negative integers (and zero), the solutions $\bar{\omega}_n$ to \eqref{eq:resonanceC} are $2\bar{\omega}_n=m-\delta-i(n+1/2)+O(\sigma^{2|\delta|})$ for non-negative integers $n$.  For the principal modes this argument no longer holds, but the QNMs turn out to take a similar form.  We quantify the error with a shift parameter $\eta$, writing \cite{Hod2008a}
\begin{align}
\label{eq:QNMfreq}
\NHN \omega_n =&  \frac{1}{2} \left[ m -\delta -  i \left(n + \frac12 \right) + \eta \right],
\end{align}
where $n$ is a non-negative integer. 
Numerical solutions of \eqref{eq:resonanceC} and direct searches for QNM frequencies at near-extremal spins show that $|\eta|$ is generally small ($\lesssim 10^{-3}$)  \cite{Yang:2013uba,Cook:2014cta}.  An analytic approximation is given in \cite{Yang:2013uba}.  Here we treat $\eta$ as a parameter and work to leading order.

Since the Green function diverges like $1/Z_{\rm in}$ near a pole, the associated residue is proportional to $\partial_\omega Z^{\rm in} = (2M/\sigma) \partial_{\bar{\omega}} Z^{\rm in}$ evaluated at $\bar{\omega}=\bar{\omega}_n$.  Using \eqref{Zin}, \eqref{eq:QNMfreq} and the expansion $1/\Gamma(-n- i \eta) = - i \eta (-1)^n n! + O(\eta^2)$, we find
\begin{align}
\label{eq:Residues}
&\left. \frac{dZ^{\rm in}}{d\NHN \omega} \right|_{\NHN \omega_n}  = 
2 \mathcal C\, \sigma^{\alpha_+}\, (-1)^n n! \Gamma(\alpha_+ - n)+ O(\eta)\,,
\end{align}
with
\begin{align}\label{C}
\mathcal C  = & -i (- i m)^{-1/2 - s + i \delta -  i m}  \\
&\times \frac{\Gamma(1 - 2 i \delta)\Gamma(-2i\delta)}{\Gamma(1/2 - s - i \delta - i m) \Gamma(1/2+s - i \delta - i m)} \,. \nonumber
\end{align}

\begin{figure}
\includegraphics[width=1.0\columnwidth]{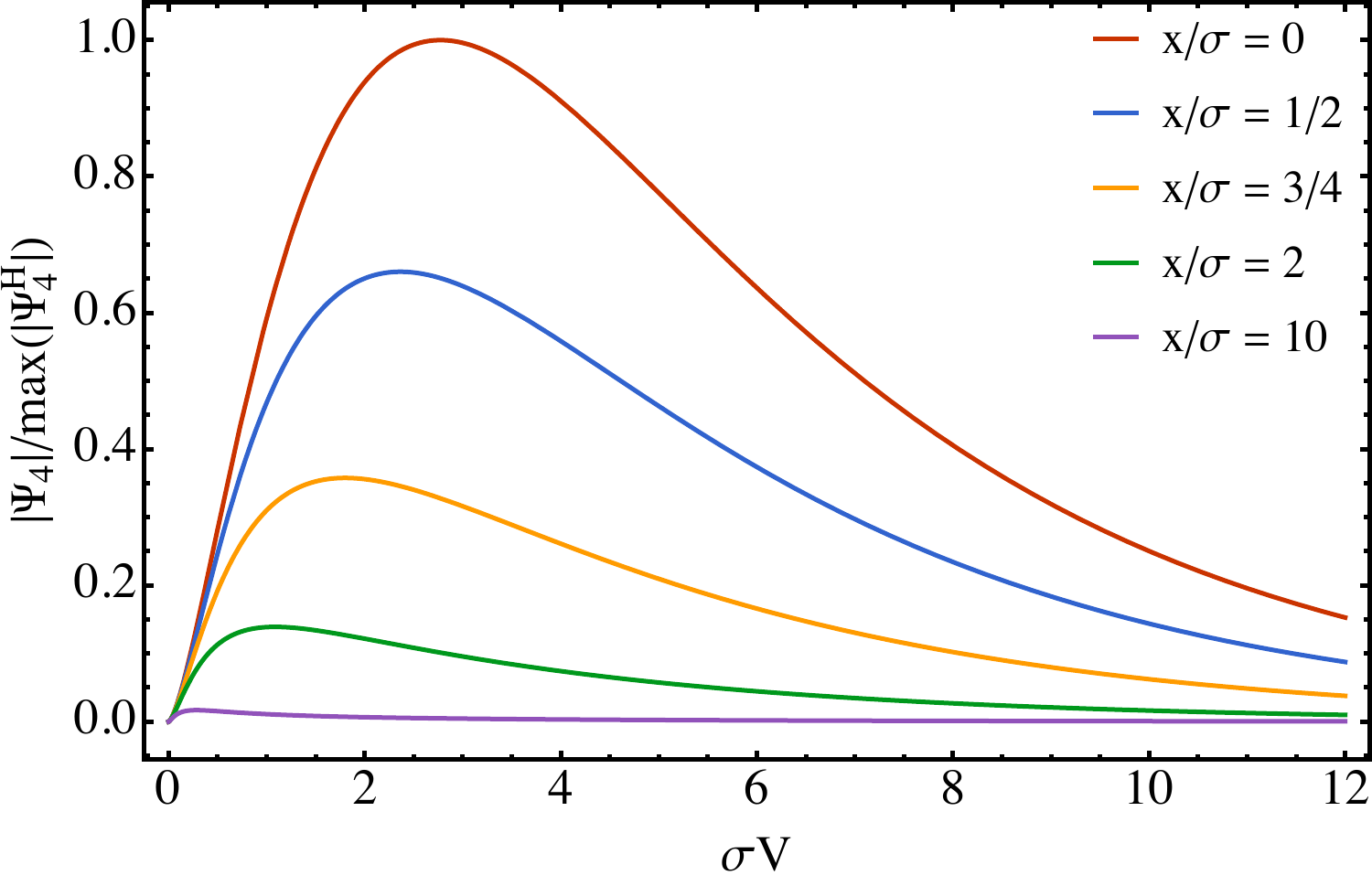} 
\caption{Plot of the magnitude of the Hartle-Hawking tetrad Weyl scalar $\Psi_4$ associated with near-horizon quasinormal modes excited by a distant pulse of $\ell = m =2$ initial data. [That is, we plot the $\ell=m=2$ term of Eq.~\eqref{eq:Gzdm} with $s=2$.]  We normalize $\Psi_4$ by the maximum value it attains on the horizon, $|\Psi_4^{\rm H}|$. The gravitational perturbations grow for a time $V \sim 1/\sigma$, and overall $\Psi_4 \sim \sigma^{-3/2}$, as determined by the scaling in Eq.~\eqref{eq:Gzdm}. }
\label{fig:GravAmp}
\end{figure}

Using the large-$r'$ form of $R_{\rm up}$ \eqref{Ruplarger} for simplicity\footnote{If $r'$ is instead any point in the far zone $(r'-r_+ \gg r_+ \sigma)$, then we have $R_{\rm up}=e^{2 i \omega r_*} f(\omega,1/r')$ for $f$ smooth near $(m/2,0)$. [See e.g.~(A5) in \cite{TeukolskyPress1974}, which can be expanded using (13.7.2) of \cite{nist}.]  The result \eqref{eq:Gzdm} is modified by replacing the $1/r'$ in front with a more complicated function of $r'$.} and dropping the $O(\eta)$ error terms, we combine Eqs.~\eqref{eq:radialg} and \eqref{eq:Residues} with our expressions for the homogeneous solutions to compute the sum over residues 
\begin{align}\label{eq:Gzdm}
G_{\rm NHM}  = &  - \frac{1} {4 r'} \sum_{\ell m} \sigma^{1-\alpha_+} \frac{ {}_sS_{\ell m}(\theta) {}_sS^*_{\ell m}(\theta')}{ m \mathcal C} 
\notag \\
& \times e^{i m(\varphi - \varphi'-V)} e^{i \delta \sigma V/2 -\sigma V/4}  \mathcal S\,,
\end{align}
where ${}_s S_{\ell m}(\theta)$ is the spin-weighted spheroidal harmonic ${}_sS_{\ell m \omega}(\theta)$  evaluated at $a=M$ and $\omega=1/(2M)$.  Here NHM stands for near-horizon modes.  We have introduced the dimensionless time coordinate 
\begin{equation} \label{V}
V = \frac{ v - v' - 2 r_*'}{2 M } \,,
\end{equation}
and the overtone sum 
\begin{align}
\label{eq:MagicSum}
\mathcal S  =
\sum_{n=0}^\infty \frac{(-1)^n e^{-n \sigma V/2} {}_2 F_1(\alpha_+, \alpha_-, \alpha_+ - n; - x/\sigma)}{n!\, \Gamma(\alpha_+ - n )} .
\end{align}
Remarkably, the sum can be evaluated in terms of elementary functions.  Taylor expanding the ${}_2F_1$ hypergeometric function, computing the sum over $n$ for each coefficient, and resumming gives
\begin{align}\label{tinyDaddy}
\mathcal S =\frac{(1 - e^{-\sigma V/2})^{\alpha_+ - 1}}{\Gamma(\alpha_+)} \left[1 + \frac{x}{\sigma}(1 - e^{-\sigma V/2} )\right]^{-\alpha_-} \,.
\end{align}
This completes the calculation of the near-horizon mode response \eqref{eq:Gzdm}.

The $\ell=m=s=2$ case is plotted in Fig.~\ref{fig:GravAmp}.  To understand the QNM response in more detail we consider the early and late time behavior.  At late times we have $\Gamma(\alpha_+) \mathcal S \to 1$, and the factor $e^{-\sigma V/4}$ in \eqref{eq:Gzdm} sets the decay rate for all modes.  At early times we have
\begin{align}\label{eq:whatababe}
 \mathcal S  &\approx  \frac{1}{\Gamma(\alpha_+)}
\left(\frac{\sigma V}{2} \right)^{\alpha_+ - 1} 
\left(1 + \frac{x V}{2} \right)^{-\alpha_-}, 
& V & \ll 1/\sigma
\,.
\end{align}
For \textit{very} early times ($V \rightarrow 0$) we cannot expect the near-horizon modes to dominate the signal, since the initial pulse of radiation arrives around $V=0$ \cite{Leaver1986b}.  The value of Eq.~\eqref{eq:whatababe} is that it reveals whether the QNM response initially grows or decays.  For $s\leq 0$ \eqref{eq:whatababe} diverges at $V=0$, corresponding to power-law decay at a rate of  $V^{-1/2+s}$.  This divergence is the usual very-early-time behavior of QNM overtone sums, which is expected to be canceled by a contribution from the branch cut \cite{Leaver1986b,Andersson:1996cm,Casals:2012tb,Casals:2012ng}.  For $s>0$ and $\delta^2>0$, however, we have the qualitatively new behavior of QNM \textit{growth} following the arrival of the signal.  

The growth lasts until a time of order $1/\sigma$, with $G_{\rm NHM}$ reaching a maximum amplitude of order $\sigma^{1/2-s}$.  From the $xV$ dependence of \eqref{eq:whatababe} we see that each higher $x$-derivative grows faster by one power of $V$.\footnote{The functional form $v^p f(xv)$ is the most general scalar that is self-similar under the NHEK dilation $v \partial_v - x \partial_x$, and hence the Aretakis behavior could have been predicted based on the principle that fields become self-similar near the horizon of extreme Kerr \cite{gralla-lupsasca-strominger2016}.} The whole approximation is valid for $x \sim \sigma$, which shrinks to the single point $x=0$ in the extremal limit.  Thus we recover the  main features of the instability: unbounded growth on the event horizon, occurring faster for higher derivatives.  The growth rates agree in detail with the extremal horizon instability \cite{Casals:2016mel,CGZsoon}.  While it should also be possible to match the full Green function (i.e.~including the numerical coefficient) in a suitable limit, the details are subtle because of the way in which all of $x \sim \sigma$ becomes compressed to $x=0$.

\section{Near-Horizon Interpretation}\label{sec:thinkOfTheMillerites}

The main result of the previous section is the portion $G_{\rm NHM}$ of the near-extremal Green function due to the near-horizon modes, which is given by Eqs.~\eqref{eq:Gzdm}, \eqref{C}, and \eqref{tinyDaddy}.  Careful inspection reveals that the answer takes the form
\begin{align}\label{essence}
G_{\rm NHM} = \sum_{\ell m} \sigma^{1/2-s -i \delta + im}\mathcal{G}_{\ell m}(\bar{x}^\mu,x^{\mu}{}'),
\end{align}
where the barred coordinates $\bar{x}^{\mu}$ are given by\footnote{We remind the reader that $V$ \eqref{V} is ingoing Kerr time in units of $2M$ and shifted to place the relevant dynamics near $V=0$.}
\begin{align}\label{barred}
\bar{x} = x/\sigma, \quad \bar{V} & = \sigma V, \quad \bar{\theta}=\theta, \quad \bar{\varphi} = \varphi - V.
\end{align}
The appearance of these coordinates is no accident: the special combinations of $x^\mu$ and $\sigma$ in \eqref{barred} are precisely what must be held fixed to produce a second regular extremal limit, the \textit{near-horizon} extremal limit which produces the NHEK metric  \cite{bardeen-horowitz1999,amseletal2009,bredbergetal2010}.

In Appendix \ref{sec:Millerites} we review these limits with the  attitude that neither is fundamentally preferred. 
The far limit ($\sigma \rightarrow 0$, fixing $x^\mu$) represents physics to  distant observers and the probes they drop into a near-extremal black hole, while the near limit ($\sigma \rightarrow 0$, fixing $\bar{x}^\mu$) represents a class of near-horizon observers and their probes. The singular relationship \eqref{barred} between the limits ensures that interactions are singular.  For example, if far probes collide with near probes, the collision energy is unbounded in the extremal limit  \cite{PSK75,Banados:2009pr,jacobson-sotiriou2009,gralla-lupsasca-strominger2016}.

The field analog of this statement is that fields smooth in one limit are singular in the other.  For example, a pulse of radiation sent towards the black hole from afar is represented by a function smooth in $x^\mu$.  But this appears highly blueshifted to near-horizon observers since $\partial_{\bar{V}} = \sigma^{-1} (2M\partial_v + \partial_\varphi)$.  Similarly, perturbations made by near-horizon observers appear to have rapid spatial variation ($\partial_x = \sigma^{-1} \partial_{\bar{x}}$) to a far-horizon probe falling into the black hole.  Of course, the distinction between space $x$ and time $V$ is artificial: in each case there are regular observers who measure arbitrarily large energies. 

Since both limits give rise to a regular limiting metric, it is natural to expect terms of the form $G(\bar{x},\bar{x}')$ and $G(x,x')$ in the near-extremal Green function, representing decoupled dynamics in the two different metrics.  
The transient instability \eqref{essence} is a kind of cross-talk $G(\bar{x},x')$, showing that far-horizon initial data can excite near-horizon modes, which are then seen as singular to infalling far-horizon observers.  
In effect, the field dynamics prevents the naive decoupling of the metrics, which manifests in the far region as an instability at small $x$.\footnote{It would be interesting to explore the reciprocal case: Does near-horizon initial data give rise to far-horizon modes that manifest in the near region as a transient instability at large $\bar{x}$?  Is there a corresponding instability of the NHEK spacetime?  This would be a linear instability, distinct from the nonlinear backreaction effects discussed in \cite{Amsel:2009ev,Dias:2009ex}.} 

From this point of view, the horizon instability of precisely extremal Kerr is nature's way of telling us that both extremal limits are always required for near-extremal perturbation theory.

\section{Physical Consequences}\label{sec:ThePhysics}

We have shown that a generic external perturbation excites a response of order $G \sim \sigma^{1/2-s}$ near the horizon ($x \sim \sigma$) at times of order $V \sim 1/\sigma$ following the initial arrival of the signal at $V=0$.  
For simplicity we imagine that some distant source acts continuously to perturb the field, so that the response is continuously of order $\sigma^{1/2-s}$.  Noting that the $x$-dependence comes only through $x/\sigma$, we may write
\begin{align}\label{lilikoi-chiffon-pie}
(\partial_x)^d G_{\rm NHM} \sim \sigma^{1/2-s-d} \quad\textrm{for} \quad x \sim \sigma.
\end{align}
For positive $s$ this response is an amplification of the external perturbation, while for any $s$ amplification occurs for sufficiently high-order derivatives.\footnote{The dependence of the growth/decay rate on $s$ can be understood in terms of the projection of the Weyl tensor onto the null tetrad~\eqref{HH}: $\Psi_4$ involves contractions onto $n^\mu$, which means that $\Psi_4$ contains directional derivatives along $n^\mu \sim (\partial_x)^\mu$, which enhance the amplitude when acting on functions of $\bar x = x/\sigma$. Meanwhile, $\Psi_0$ contains directional derivatives along the direction $l^\mu \sim (\partial_v)^\mu + \sigma x M^{-1} (\partial_x)^\mu + (2M)^{-1}(\partial_{\phi})^\mu$, and these derivatives do not provide enhancements when acting on functions of $\bar x^\mu$.}

In the scalar case $s=0$ the Green function $G$ refers to a massless scalar field $\Phi$ propagating on the Kerr background.\footnote{Note that if the factors of $m$ are replaced with $2 r_+ \omega$ in \eqref{alpha}, \eqref{delta2} and \eqref{eq:Residues}, and we express $\omega$ in terms of $\Omega_H$ and $\bar \omega$, our results carry over to scalar fields in near-extremal Kerr-Newman backgrounds (but not for electromagnetic or gravitational fields). See e.g.~\cite{Zimmerman:2015trm}.}  Thus the field values ($d=0$) are modest ($\Phi \sim \sigma^{1/2}$), but the first derivative becomes large ($\partial_x \Phi \sim \sigma^{-1/2}$).  The stress-energy tensor $T_{\mu \nu}$ is quadratic in first derivatives, and infalling observers $u^\mu$ generically see large energy densities,\footnote{Equation \eqref{energetic} also holds for generic observers near the black hole ($x \sim \sigma$) whose four-velocity has a smooth far-horizon limit, but we specifically think of observers dropped from a large radius (without any fine-tuning).  These observers pass through all values of $x$ and hence experience large energy densities at some period in their journey.}
\begin{align}\label{energetic}
E_{\rm obs} = T_{\mu \nu} u^\mu u^\nu \sim \sigma^{-1} \rightarrow \infty.
\end{align}
This is analogous to the high-energy particle collisions \cite{Banados:2009pr} that can occur in the near-horizon region with sufficient fine-tuning (see  Appendix \ref{sec:Millerites}). 
Here, on the other hand, any generic external perturbation excites near-horizon modes so that a generic particle sent in experiences a high-energy ``collision'' \eqref{energetic} with the field.

If an infalling observer carried some scalar charge, then in addition to large energies \eqref{energetic} she would also experience large forces $\partial_x \Phi \sim 1/\sqrt{\sigma}$.  Of course, she may pass through the small region $x \sim \sigma$ too quickly to notice any significant change in her trajectory.  Similarly, Eq.~\eqref{energetic} represents an energy \textit{density}, and the effect on an observer over the region $x \sim \sigma$ may in fact be finite.  Resolving these questions would require a definite calculation within some scalar model.

In the electromagnetic case $s=\pm 1$, the Green function $G$ corresponds to ingoing Kerr components of the field strength tensor $F_{\mu \nu}$ (see Appendix \ref{sec:tetrad} for details).  The growing case $s=1$ corresponds by Eq.~\eqref{Omega1} to the Hartle-Hawking scalar $\phi_2$, which contains $F_{rv}$, $F_{r\theta}$, and $F_{r \varphi}$.  By \eqref{lilikoi-chiffon-pie} we have $\phi_2 \sim 1/\sqrt{\sigma}$.  The stress-energy is quadratic in $F$ so again large energies \eqref{energetic} are generically observed by infalling observers.  For extremely rapidly spinning black holes this could in principle allow an astrophysical probe of high-field quantum electrodynamics by infalling charged particles.

More likely to have an interesting astrophysical effect are the large Lorentz forces $F_{\mu \nu} u^\nu \sim 1/\sqrt{\sigma}$.  In effect, rapidly spinning black holes amplify external electromagnetic perturbations by a factor of $1/\sqrt{\sigma}$.  Free charges moving toward the black hole would have their bulk motion and synchrotron spectra suitably modified by the enhanced field near the horizon.  This provides a promising avenue for astrophysical signatures of the instability, especially if coupled with a transient behavior while the field ramps up over times of order $1/\sigma$.  However, while we can expect distinctive features near the horizon, these may be washed out as the radiation climbs out of the gravitational potential well.  Detailed calculation is required to determine a precise astrophysical signature.

In the gravitational case, $s=\pm 2$, the Green function $G$ refers to ingoing Kerr components of the Weyl tensor $C_{\mu \nu \rho \sigma}$ (see Appendix \ref{sec:tetrad} for details).  The growing case $s=2$ corresponds by Eq.~\eqref{Omega2} to the Hartle-Hawking scalar $\Psi_4$, which contains components with two appearances  of $r$ (e.g.~$C_{vrvr}$).  By \eqref{lilikoi-chiffon-pie} we have $\Psi_4 \sim \sigma^{-3/2}$.  This represents a relative enhancement of the tidal forces felt by an infalling observer compared to what she would feel near a comparable modestly spinning black hole.  However, for astrophysically  reasonable parameters the forces would be swamped by those of the black hole itself.

A more promising route to an astrophysical signature of the transient gravitational instability is through its contribution to a nonlinear parametric resonance that may drive gravitational turbulence \cite{Yang:2014tla}.  This resonance occurs because the near-horizon modes have approximately integer-separated frequencies from the far-zone perspective, $\omega_{\rm NHM}=m/2+O(\sigma)$.  The authors of Ref.~\cite{Yang:2014tla} calculated a criterion for the onset of turbulence assuming that the driving perturbation $h$ is due to the single lowest overtone.
Our results indicate that coherent excitation gives rise to power-law decay or growth of near-zone perturbations.  Accounting for this could modify the criterion for the onset of turbulence, likely enhancing the effect.

\section*{Acknowledgements}

This work was supported in part by NSF Grant No.~PHY--1506027 to the University of Arizona and in part by Perimeter Institute for Theoretical Physics. Research at the Perimeter Institute is supported by the Government of Canada through Industry Canada and by the Province of Ontario through the Ministry of Economic Development \& Innovation.

\appendix

\section{Tetrad}\label{sec:tetrad}
The field variable we employ, $\Omega_s$,  is defined in a time-reversed version of the Kinnersley tetrad which is regular on the future horizon. A more common tetrad, which is also regular on the horizon, is the Hartle-Hawking (HH) tetrad \cite{TeukolskyPress1974}. Here we relate $\Omega_s$ to physical quantities in the HH tetrad.

The HH tetrad is obtained from the Kinnersley tetrad by the type-III null
transformation  $\ell^\mu \to \Lambda \ell^\mu$, $n^\mu \to \Lambda^{-1}n^\mu$, $ m^\mu \to e^{ i \chi} m^\mu$, with $\chi=0$ and boost parameter $\Lambda = \Delta/[2 (r^2+a^2)]$.   The HH tetrad  has ingoing Kerr components
\begin{align}\label{HH}
\ell^\mu & = \left(1,\frac{\Delta}{2(r^2+a^2)},0,\frac{a}{r^2+a^2}\right), \\
n^\mu & = \left(0,-\frac{r^2+a^2}{r^2 + a^2 \cos^2 \theta },0,0\right), \\
m^\mu & = \frac{1}{\sqrt{2}(r+i a \cos \theta)}\left(i a \sin \theta,0,1,\frac{i}{\sin \theta}\right).
\end{align}
The leg $\ell^\mu$ is tangent to the horizon, while $n^\mu$ is transverse.

The scalar $\Omega_s$ is related to the HH field scalar, $\Upsilon^{\rm HH}_s$, by $\Omega_s = (r^2+a^2)^s \Upsilon^{\rm HH}_{-s}$ \cite{TeukolskyPress1974}. 
The electromagnetic scalars $\Omega_{\pm 1 }$, which correspond to the $|s|=1$ Green function derived in \ref{sec:GreenFunc}, are related to the HH scalars $\phi_0 = F_{\alpha \beta } \ell^{\alpha} m^\beta$ and $\phi_2 = F_{\alpha \beta } \bar m^\alpha n^\beta$ via
\begin{align}
\Omega_{-1} &= (r^2 + a^2)^{-1}  \phi_0, \label{Omegam1}  \\
 \Omega_{1} &= (r^2+a^2)(r-i a \cos \theta)^{2}  \phi_2, \label{Omega1}
\end{align}
where the overbar indicates complex conjugation. The remaining components, $\phi_1 = \frac12 F_{\alpha \beta} \left(\ell^\alpha n^\beta + \bar m^\alpha m^\beta \right)$ may be obtained either by solving 
a first-order partial differential equation \cite{Teukolsky1973}, or by performing a field reconstruction such as that outlined in \cite{Cohen:1974cm}. 
Similarly, the gravitational  scalars $\Omega_{\pm 2}$ derived in the text are related to the radiative components of the gravitational field, $\Psi_4 =  C_{\alpha \mu \beta \nu} n^\alpha \bar m^\mu n^\beta \bar m^\nu$ and $\Psi_0 =  C_{\alpha \mu \beta \nu} \ell^\alpha  m^\mu \ell^\beta  m^\nu$,
via
\begin{align}
 \Omega_{-2} &= (r^2 + a^2)^{-2}  \Psi_0, \label{Omegam2} \\ 
 \Omega_2  &= (r^2+a^2)^{2} (r-i a \cos \theta)^{4}\Psi_4. \label{Omega2}
\end{align}

\section{Extremal limits}\label{sec:Millerites}
The near-horizon coordinates \eqref{barred} that capture the essential properties of the instability have a rather mysterious origin in the calculations of \cite{TeukolskyPress1974} and this paper.  We now give a discussion of near-extremal physics that leads naturally to these coordinates and their associated near-horizon extremal limit. 
As in the text, we use ingoing Kerr coordinates $x^\mu=(v,r,\theta,\varphi)$, together with the useful definitions [repeated from \eqref{sigmax}]
\begin{align}\label{sigmax2}
\sigma = \frac{r_+-r_-}{r_+}, \qquad x = \frac{r-r_+}{r_+}.
\end{align}

We begin with an analysis of equatorial orbits.  For any nonextremal Kerr black hole there are three particularly interesting prograde circular orbits \cite{bardeen-press-teukolsky1972}, located to leading order in $\sigma$ at
\begin{subequations}\label{three-musketeers}
\begin{align}
x_{\rm ISCO} & = 2^{1/3} \sigma^{2/3} \label{xISCO} \\
x_{\rm IBCO} & = (\sqrt{2}-1) \sigma \label{xIBCO} \\
x_{\rm ICO} & = (2/\sqrt{3}-1) \sigma. \label{xICO}
\end{align}\end{subequations}
The innermost stable circular orbit (ISCO) is a marginally stable orbit separating the stable orbits at larger radii from the unstable orbits at smaller radii.  The innermost bound circular orbit (IBCO) similarly separates bound orbits from unbound orbits.\footnote{By bound we mean with ratio of energy to rest mass less than unity.  The unbound, unstable circular orbits have the property that small perturbations directed outward cause the particle to escape to infinity instead of settling into a bound orbit.}  The innermost circular orbit (ICO) is a null orbit inside of which there are no circular orbits at all.  Note that $x_{\rm ICO}<x_{\rm IBCO}<x_{\rm ISCO}$ for all $\sigma>0$.

These orbits are important for various physical processes in Kerr.  For example, accretion disks terminate somewhere between the ISCO and the IBCO \cite{abramowicz-fragile2013}, depending on the thickness of the disk.  A near-equatorial compact object inspiraling into the black hole (a promising source of gravitational radiation \cite{eLISA}) would similarly end its journey by orbiting many times in this region before plunging in \cite{glampedakis-kennefick2002}.  The ICO is important for photon propagation, determining, among other things, the size and shape of the shadow \cite{bardeen1973} cast by a black hole, which the Event Horizon Telescope \cite{EHT} hopes to measure.

The need for a second extremal limit can be seen from the way the standard (far zone) one completely muddles these important orbits, making them all coincide [\eqref{three-musketeers} as $\sigma \rightarrow 0$].  In fact the situation is worse, since they approach the horizon $x=0$ of extremal Kerr and hence become \textit{null}.  This manifests as a blowing up of the four-velocity of the timelike orbits.  In particular, for the IBCO we have
\begin{align}\label{uIBCO}
u_{\rm IBCO} =  \frac{\sqrt{8}}{\sigma} \left( \partial_v + \frac{1}{2M} \partial_\varphi \right) - \left( \partial_v + \frac{2}{M} \partial_\varphi \right) + O(\sigma).
\end{align}
Every circular orbit inside the ISCO suffers a similar fate.  
Clearly, the usual extremal limit drastically distorts the near-horizon physics.

To preserve the near-horizon physics we should take a different limit where the critical orbits stay distinct.  It is clear from \eqref{three-musketeers} that the IBCO and ICO stay at finite coordinate radius if we use $x/\sigma$ instead of $x$.  To preserve the timelike character of the IBCO, we must stop the blowup in \eqref{uIBCO} by finding new time and angular coordinates such that $(\partial_v + (2M)^{-1} \partial_\varphi)/\sigma$ is finite.  This can be accomplished by rescaling $v$ and shifting $\varphi$, making the complete set\footnote{These agree with the barred coordinates \eqref{barred} used in the text up to an irrelevant shift in the origin of time in \eqref{V}.}
\begin{align}\label{barred2}
\bar{v} = \frac{\sigma v}{2M}, \quad \bar{x} = \frac{x}{\sigma}, \quad \bar{\theta}=\theta, \quad \bar{\varphi} = \varphi - \frac{v}{2M}.
\end{align}
If we let $\sigma \rightarrow 0$ fixing barred coordinates $\bar{x}^\mu$ then the IBCO remains timelike and distinct from the horizon,
\begin{align}
\bar{x}_{\rm IBCO} = \sqrt{2}-1, \quad u_{\rm IBCO} = \frac{\sqrt{2}}{M} \left( \partial_{\bar{v}} - \frac{3}{2\sqrt{2}} \partial_{\bar{\varphi}} \right).
\end{align}
Here we have kept to leading order in $\sigma$ at fixed $\bar{x}^\mu$.  Having been led to the rather nontrivial scalings in \eqref{barred2} by considering the IBCO, and we may now check that these coordinates provide a good limit for the entire metric as well.  Letting $\sigma \rightarrow 0$ fixing $\bar{x}^\mu$ in the Kerr metric yields
\begin{align}\label{NHEK}
ds^2 & = 2 M^2 \Gamma(\theta)\big[ -\bar{x}(\bar{x}+2) d\bar{v}^2 +2 d\bar{v} d\bar{x} + d\theta^2 \nonumber \\ & \qquad \qquad + \Lambda(\theta)^2 \left( d\bar{\varphi} + (\bar{x}+1)d\bar{v} \right)^2 \big]
\end{align}
where $\Gamma(\theta) = (1+\cos^2 \theta)/2$ and $\Lambda(\theta)=2\sin \theta/(1+\cos^2\theta)$.  This is the NHEK metric in coordinates adapted to the future horizon of near-extremal Kerr.\footnote{In some versions of the near-horizon limit of near-extreme Kerr one introduces a scaling parameter $\lambda$ and lets $\sigma = \lambda \bar{\sigma}$.  Then the $\lambda \rightarrow 0$ limit produces the metric \eqref{NHEK} with the numerals $2$ and $1$ replaced by $2\bar{\sigma}$ and $\bar{\sigma}$, respectively.}  It has a number of interesting properties, notably two extra Killing fields that enhance the isometry group to $\mathsf{SL}(2,\mathbb{R})\times\mathsf U(1)$ \cite{bardeen-horowitz1999}.  Here we only point out that it is not asymptotically flat: the far-horizon region has disappeared.  Thus the situation is rather symmetric, with each limit faithful to one region but not the other.

The IBCO has been our muse, but any timelike curve in NHEK  represents the experience of some physical observer near a rapidly rotating black hole.  Formally, we may represent an observer in near-extremal Kerr by a family of timelike orbits, each defined on a separate nonextremal Kerr spacetime, parametrized by $\sigma$.  We call orbits with a good near-horizon limit (four-velocity finite and nonzero) near-horizon observers, while those with a good far-horizon limit are called far-horizon observers.  The physical question at hand determines which observers to consider, but we see no fundamental reason to prefer either.

An important observation is that the two kinds of observers are at infinite relative boost in the limit \cite{gralla-lupsasca-strominger2016}. 
This is evident from the singular relationship \eqref{barred2} between the limits.  A simple example is the IBCO (a near-horizon observer) and a generic infalling observer.  From \eqref{uIBCO} we see that the boost factor $u^\alpha_{\rm IBCO} u_\alpha$ with some far-horizon observer $u$ diverges like $\sigma^{-1}$ except in the fine-tuned case $u_\varphi=2 M u_v + O(\sigma)$.  The black hole can be regarded as a ``particle accelerator'' if instead of placing the first particle on the IBCO, one drops it in from infinity with precisely the right parameters so that it asymptotically orbits on the IBCO \cite{Banados:2009pr,jacobson-sotiriou2009}.  A second particle dropped in later then collides at high energy.  In this way we can view the existence of high-energy collisions as a consequence of the existence of two limits at infinite relative boost \cite{gralla-lupsasca-strominger2016}. 

The field analog of this statement is that fields with a smooth near limit look singular in the far limit, and vice versa.  As described in detail in  Sec.~\ref{sec:thinkOfTheMillerites}, the relationship \eqref{barred2} between the two limits ensures that if fields are smooth in one limit, then  sufficiently high-order derivatives blow up in the other.  Thus one can ensure singular behavior simply by considering a source or initial data adapted to one limit and an observer adapted to the other.
The instability discussed here is the further statement that in fact one \textit{cannot avoid} singular behavior by avoiding near-horizon sources, since generic far-horizon perturbations excite near-horizon modes.

For completeness we now discuss the ISCO.  From \eqref{xISCO} we see that this orbit scales as $\sigma^{2/3}$ and therefore is irregular in both the near limit ($\bar{x}_{\rm ISCO}\rightarrow \infty$) \textit{and} the far limit (where $x \rightarrow 0$).  One can take a third limit adapted to the ISCO scaling, which produces a different coordinate patch of the NHEK spacetime \cite{hadar-porfyriadis-strominger2014,GPW15}, but this limit is not particularly useful as it covers neither the horizon nor the asymptotic region.  It seems most useful to regard marginally stable geodesics as living at a very large radius in the near-horizon metric, much as we would regard stationary geodesics as living at a very large radius in the far-horizon metric.

\bibliographystyle{apsrev4-1}
\bibliography{RefsMaster}

\end{document}